\documentclass{PoS}
\usepackage{cite}

\title{N$^3$LO calculations for top-quark differential cross sections near partonic threshold}

\ShortTitle{N$^3$LO calculations for top-quark differential cross sections near partonic threshold}

\author{\speaker{Nikolaos Kidonakis}\thanks{This material is based upon work supported by the National Science Foundation under Grant No. PHY 1212472 and PHY 1519606.}\\
   Department of Physics, Kennesaw State University, Kennesaw, GA 30144, USA\\
        E-mail: \email{nkidonak@kennesaw.edu}}

\abstract{I present calculations of approximate corrections from NNLL soft-gluon resummation for total and differential cross sections in top-antitop pair production and single-top production in hadronic collisions. I show that soft-gluon corrections from partonic threshold are the dominant contribution to top-quark production and closely approximate exact results. I present aN$^3$LO results for the total $t{\bar t}$ cross section, the top-quark $p_T$ and rapidity distributions, and the top-quark forward-backward asymmetry. I also provide updates for single-top production in the $t$, $s$, and $tW$ channels. Finally, I discuss top-quark production via anomalous gluon couplings.}

\FullConference{Radcor 2015 and Loopfest XIV\\
		15-19 June 2015\\
                Los Angeles, CA, USA}

\def\beq{\begin{equation}}
\def\eeq{\end{equation}}
\def\beqa{\begin{eqnarray}}
\def\eeqa{\end{eqnarray}}

\begin{document}

\section{Introduction}

Accurate and precise calculations of higher-order perturbative corrections for $t{\bar t}$ total cross sections, top-quark transverse momentum ($p_ T$) distributions, 
top-quark rapidity distributions and forward-backward asymmetry ($A_{\rm FB}$), 
as well as for single-top production cross sections and differential 
distributions, are of primary importance for top-quark physics.
Higher-order QCD corrections are very significant for top-antitop pair 
production and for single-top production.
Soft-gluon corrections are large, particularly near threshold, and they are the dominant part of the QCD corrections at LHC and Tevatron energies. 

The partonic threshold approximation used in the calculation of soft-gluon corrections in our formalism \cite{NK2009,NK2010,NK2011} is more general than absolute production threshold in that the top quarks are not produced at rest. Moreover, the method is applied to double-differential cross sections and used to derive total cross sections and differential distributions. It has been known for some time that the partonic threshold approximation in our approach works very well for LHC and Tevatron energies; for $t{\bar t}$ production, the difference between our approximate \cite{NK2010,NK2011} and the exact cross sections at both 
NLO \cite{NLO1,NLO2,NLO3} and NNLO \cite{CFM,CFM2} are at the per mille level. 
These approximate soft-gluon corrections are currently known through N$^3$LO \cite{NKaNNNLO,NKpty,NKafb}.

The perturbative QCD corrections include soft-gluon terms of the form  
$[\ln^k(s_4/m_t^2)]/s_4$, where $k \le 2n-1$ at $n$th order, 
and $s_4$ is the kinematical distance from partonic threshold.
These soft corrections are resummed at next-to-next-to-leading logarithm (NNLL) accuracy via factorization and 
renormalization-group evolution of soft-gluon functions through two loops 
\cite{NK2009,NK2010}. Resummation is performed on the double-differential cross section 
using the standard moment-space formalism, and the cross section is then expanded to 
fixed order. The use of a fixed-order expansion removes the need for prescriptions (such as the minimal prescription) to deal with divergences, and thus it avoids the unphysical effects of such prescriptions, as explained in \cite{NK2000}. 
In particular, our approximate N$^3$LO (aN$^3$LO) results differ from and are more accurate than the NNLO+NNLL results of \cite{CFM}, in part because we do not use the minimal prescription and in part because we employ a more general resummation formalism at the differential level using partonic threshold rather than absolute threshold. 

The first N$^3$LO expansion for cross sections was given in \cite{NK2000}, and a complete general expression was given in \cite{NK2005}.
The latest aN$^3$LO results from NNLL resummation for the $t{\bar t}$ total cross section \cite{NKaNNNLO}, top $p_T$ and rapidity distributions \cite{NKpty}, and the top forward-backward asymmetry $A_{\rm FB}$ \cite{NKafb}, provide the most accurate and precise theoretical predictions to date. 
The stability and reliability of the theoretical higher-order results in our
resummation approach over the past two decades as well as the correct prediction of the size of the exact NNLO corrections fully validate our formalism.
The agreement with top-pair LHC and Tevatron data \cite{ttbar7lhccombo,ttbar8lhccombo,ttbartevcombo,CMStoppty8lhc,CDFafb,D0afb} is excellent.

For single-top production, approximate NNLO (aNNLO) total and differential cross sections have been derived from the expansion of the NNLL resummed expressions \cite{NKtch,NKsch,NKtW,NKppn,NKtchpt} in all three channels. Again, there is excellent agreement with LHC and Tevatron data \cite{tchtevcombo,ATLAStch7lhc,CMStch7lhc,ATLAStW7lhc,CMStW7lhc,tch8lhccombo,ATLASsch8lhc,CMSsch8lhc,tW8lhccombo}.

In the following we present first theoretical results for top-antitop pair production followed by results for single-top production. In both cases we compare with recent experimental data and find excellent agreement between theory and experiment. We also briefly discuss top production via anomalous gluon couplings.

\section{$t{\bar t}$ production}

We begin with $t{\bar t}$ production, and we present results for total cross sections and differential distributions as well as the forward-backward asymmetry. We use MSTW2008 NNLO pdf \cite{MSTW2008} for all our predictions.

\subsection{$t{\bar t}$ total cross sections at the LHC and the Tevatron}

We first study the total cross section through aN$^3$LO \cite{NKaNNNLO}. This means that the approximate N$^3$LO soft-gluon corrections are added to the exact NNLO result to provide the best prediction.

\begin{figure}
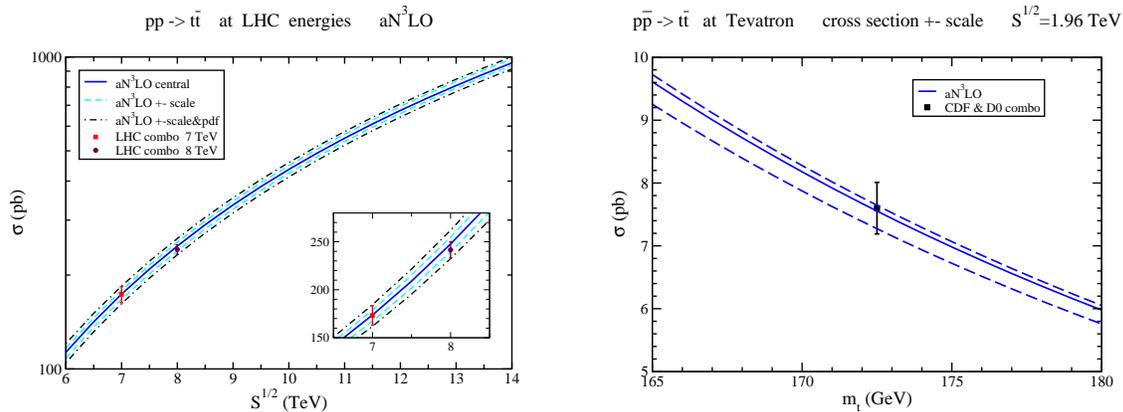

\begin{center}
\includegraphics[width=.45\textwidth]{ttSlhcaN3LOplot.eps}
\hspace{10mm}
\includegraphics[width=.45\textwidth]{tevatronaN3LOnplot.eps}
\caption{Total aN$^3$LO cross sections for $t{\bar t}$ production at the LHC (left) and the Tevatron (right).}
\label{figttbar}
\end{center}
\end{figure}

In Fig. \ref{figttbar} we display the aN$^3$LO total $t{\bar t}$ cross sections 
at the LHC versus collider energy, and at the Tevatron versus top-quark mass. We compare our results with LHC combination data from the ATLAS and CMS collaborations at 7 TeV \cite{ttbar7lhccombo} and 8 TeV \cite{ttbar8lhccombo} energies, and with Tevatron combination data from the CDF and D0 collaborations at 1.96 TeV energy \cite{ttbartevcombo}. We find excellent agreement of the theoretical predictions with the experimental data in all cases. 

\begin{table}[htb]
\begin{center}
\begin{tabular}{c|c|c|c|c|c}
${\sqrt S}$ & 1.96 TeV  & 7 TeV & 8 TeV & 13 TeV & 14 TeV \\ 
\hline $\sigma$ 
& $7.37 {}^{+0.09}_{-0.27} {}^{+0.38}_{-0.28}$
& $174 {}^{+5}_{-7}  {}^{+9}_{-10}$
& $248 {}^{+7}_{-8}  {}^{+12}_{-13}$
& $810 {}^{+24}_{-16}{}^{+30}_{-32}$
& $957 {}^{+28}_{-19}{}^{+34}_{-36}$
\end{tabular}
\caption{aN$^3$LO $t{\bar t}$ total cross sections in pb with $m_t=173.3$ GeV. The 1.96 TeV energy is for Tevatron $p{\bar p}$ collisions while the 7, 8, 13, and 14 TeV energies are for LHC $pp$ collisions.}
\label{table1}
\end{center}
\end{table}

In Table \ref{table1} we show the values of the aN$^3$LO total $t{\bar t}$ cross sections in pb for a top-quark mass $m_t=173.3$ GeV at Tevatron and LHC energies. The central results are for scale $\mu=m_t$. The first uncertainty in the cross sections is from scale variation over the range $m_t/2 \le \mu \le 2m_t$, and the second is from the MSTW2008 pdf \cite{MSTW2008} at 90\% C.L.

It is quite interesting to study the convergence of the perturbative series for the cross section.
We write the series through aN$^3$LO as 
$\sigma^{\rm aN^3LO}=\sigma^{(0)} \left[1+\frac{\sigma^{(1)}}{\sigma^{(0)}}
+\frac{\sigma^{(2)}}{\sigma^{(0)}}+\frac{\sigma^{(3)}}{\sigma^{(0)}}\right]$, 
where $\sigma^{(0)}$ is the LO cross section, $\sigma^{(1)}$ denotes 
the complete NLO corrections, $\sigma^{(2)}$ denotes the complete NNLO 
corrections, and $\sigma^{(3)}$ denotes the approximate N$^3$LO soft-gluon 
corrections. The values for the fractions $\sigma^{(n)}/\sigma^{(0)}$, 
with $n=1,2,3$, are shown in Table \ref{table2}.

\begin{table}[htb]
\begin{center}
\begin{tabular}{|c|c|c|c|c|c|} \hline
\multicolumn{6}{|c|}{Perturbative series through aN$^3$LO for the $t{\bar t}$ cross section} 
\\ \hline
corrections & Tevatron & LHC 7 TeV & LHC 8 TeV & LHC 13 TeV& LHC 14 TeV \\ \hline
$\sigma^{(1)}/\sigma^{(0)}$ & 0.236 & 0.470 & 0.476 & 0.493 & 0.496 \\ \hline 
$\sigma^{(2)}/\sigma^{(0)}$ & 0.106 & 0.178 & 0.177 & 0.172 & 0.170 \\ \hline 
$\sigma^{(3)}/\sigma^{(0)}$ & 0.068 & 0.066 & 0.059 & 0.045 & 0.043 \\ \hline 
\end{tabular}
\caption[]{The fractional contributions at higher orders relative to LO, 
all calculated with the same 
NNLO pdf \cite{MSTW2008}, to the $t{\bar t}$ production 
cross section at the Tevatron 
with $\sqrt{S}=1.96$ TeV and at the LHC with 
$\sqrt{S}=7$, 8, 13, and 14 TeV, with $\mu=m_t=173.3$ GeV.}
\label{table2}
\end{center}
\end{table}

The fractional contributions to the perturbative series at LHC energies converge well through aN$^3$LO, which could potentially indicate that corrections beyond N$^3$LO are negligible \cite{NKaNNNLO}. At 14 TeV, $\sigma^{\rm aN^3LO}=1.709 \, \sigma^{(0)}$ as can be found by summing the corresponding entries in Table \ref{table2}. The series $\sum_{n=1}^4 1/n!=1.708\cdots$ approximates the situation very well. If this trend would continue in higher orders, which is of course not known, then using the result  $\sum_{n=1}^{\infty} 1/n!=e-1=1.718\cdots$ one would expect no more than 0.5\% increase of the aN$^3$LO cross section from  additional corrections beyond N$^3$LO. The above remarks describe the 13 TeV results very well too.

\subsection{Top-quark $p_T$ and rapidity distributions at the LHC and the Tevatron}

We continue with the top-quark transverse momentum and rapidity distributions in $t{\bar t}$ production at the LHC and the Tevatron.

\begin{figure}
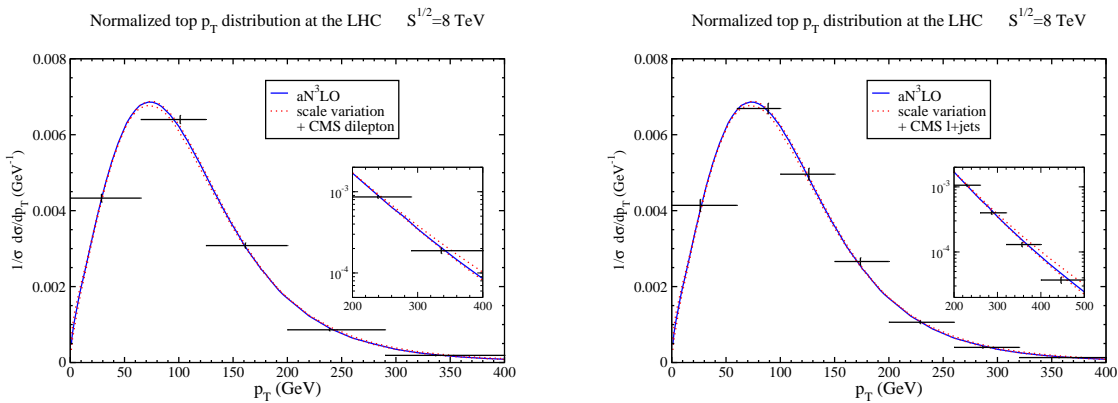

\begin{center}
\includegraphics[width=.45\textwidth]{pt8lhcnormCMSdileptplot.eps}
\hspace{10mm}
\includegraphics[width=.45\textwidth]{pt8lhcnormCMSleptjetplot.eps}
\caption{Normalized aN$^3$LO top-quark $p_T$ distributions at the 8 TeV LHC and comparison with CMS data in the dilepton (left) and lepton+jets (right) channels.}
\label{figpt8lhc}
\end{center}
\end{figure}

In Fig. \ref{figpt8lhc} we show the normalized aN$^3$LO top-quark $p_T$ distribution, $(1/\sigma) d\sigma/dp_T$, at 8 TeV LHC energy and compare with recent results from CMS \cite{CMStoppty8lhc} separately in the dilepton and lepton+jets channels. The theoretical predictions are in excellent agreement with the data. 
As shown in \cite{NK15tt}, the theory also is in excellent agreement with 7 TeV LHC data and 1.96 TeV Tevatron data.

\begin{figure}
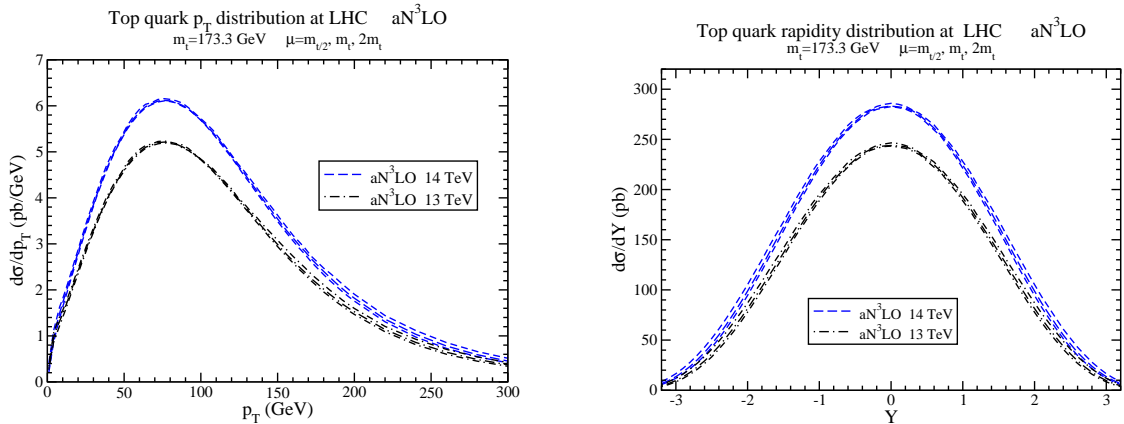

\begin{center}
\includegraphics[width=.45\textwidth]{pt13and14lhcaN3LOnplot.eps}
\hspace{10mm}
\includegraphics[width=.45\textwidth]{y13and14lhcaN3LOnplot.eps}
\caption{Top-quark aN$^3$LO $p_T$ distributions (left) and rapidity distributions (right) at 13 and 14 TeV LHC energies.}
\label{figptylhc}
\end{center}
\end{figure}

In the left plot of Fig. \ref{figptylhc} we show the aN$^3$LO top-quark 
$p_T$ distributions, $d\sigma/dp_T$, at 13 and 14 TeV LHC energies.

We continue with the top quark rapidity distribution at the LHC.
In the right plot of Fig. \ref{figptylhc} we show the aN$^3$LO  
rapidity distributions, $d\sigma/dY$, at 13 and 14 TeV LHC energies.

\begin{figure}
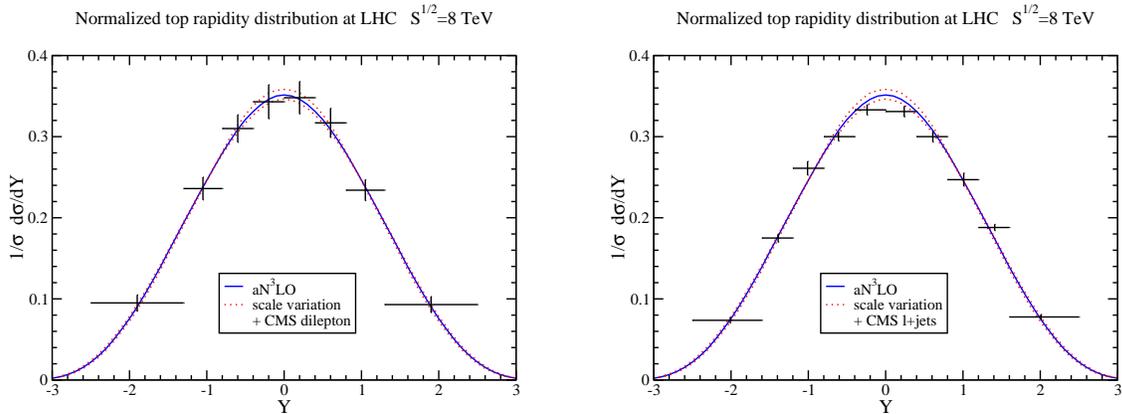

\begin{center}
\includegraphics[width=.45\textwidth]{y8lhcnormCMSdileptplot.eps}
\hspace{10mm}
\includegraphics[width=.45\textwidth]{y8lhcnormCMSleptjetplot.eps}
\caption{Top-quark aN$^3$LO normalized rapidity distributions at the 8 TeV LHC and comparison with CMS data in the dilepton (left) and lepton+jets (right) channels.}
\label{figy8lhc}
\end{center}
\end{figure}

In Fig. \ref{figy8lhc} we show the normalized aN$^3$LO top-quark rapidity distribution, $(1/\sigma) d\sigma/dY$, at 8 TeV LHC energy and compare with recent results from CMS in the dilepton and lepton+jets channels \cite{CMStoppty8lhc}. Again, theory and data agree very well. As shown in \cite{NK15tt}, the theoretical predictions are also in excellent agreement with 7 TeV LHC data and 1.96 TeV Tevatron data.

\subsection{Top-quark forward-backward asymmetry at the Tevatron}

Next, we present results for the top-quark forward-backward asymmetry at the Tevatron. As was discussed in \cite{NKafb}, the soft-gluon corrections are dominant and our formalism precisely predicted in Ref. \cite{NK2011} the increase in the exact asymmetry at NNLO in \cite{CFM2}. The aN$^3$LO perturbative corrections calculated in \cite{NKafb} are large and they improve the NNLO result of \cite{CFM2}. Including electroweak corrections and the aN$^3$LO QCD corrections, we find an asymmetry of ($10.0 \pm 0.6$)\% in the $t{\bar t}$ frame \cite{NKafb}. 

The differential top forward-backward asymmetry is defined by 
\beqa
A^{\rm bin}_{\rm FB} &=&\frac{\sigma^+_{\rm bin}(\Delta y)-\sigma^-_{\rm bin}(\Delta y)}
{\sigma^+_{\rm bin}(\Delta y)+\sigma^-_{\rm bin}(\Delta y)} \hspace{5mm}
{\rm with} \hspace{3mm} \Delta y=y_t-y_{\bar t} \, .
\label{AFBbin}
\nonumber
\eeqa

\begin{figure}
\begin{center}
\includegraphics[width=.65\textwidth]{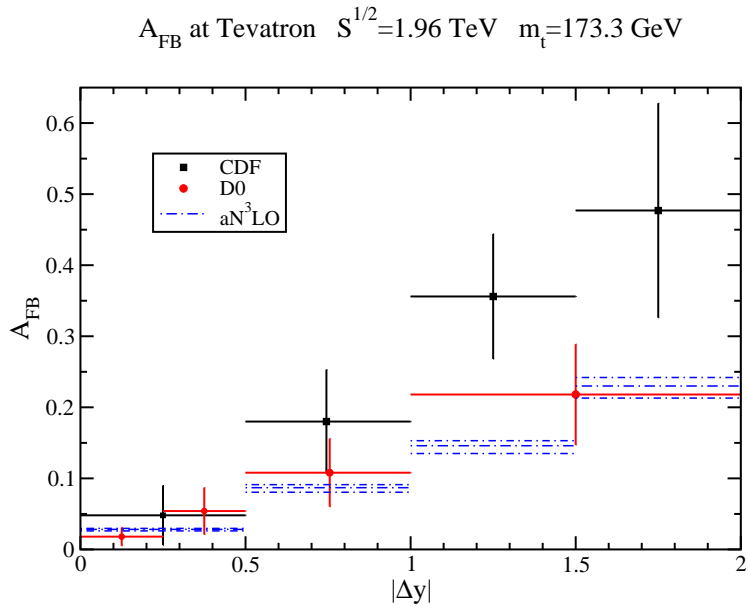}
\caption{Top-quark aN$^3$LO differential $A_{\rm FB}$ at the Tevatron.}
\label{figytev}
\end{center}
\end{figure}

In Fig. \ref{figytev} we plot the differential $A_{\rm FB}$, including QCD corrections through aN$^3$LO \cite{NKafb}, and compare with recent results from CDF \cite{CDFafb} and D0 \cite{D0afb}. The agreement between theory and data is very good for both the total and the differential asymmetries.

\section{Single-top production}

Single-top production can be categorized into three channels: the $t$-channel, the $s$-channel, and the $tW$ channel. NNLL resummation for all these processes was performed in Refs. \cite{NKtch,NKsch,NKtW,NKppn,NKtchpt}. We add approximate NNLO (aNNLO) soft-gluon corrections to the exact NLO results \cite{NLOst1,NLOst2} for our best predictions.

\begin{table}[htb]
\begin{center}
\begin{tabular}{c|c|c|c}
LHC  & $t$-channel & $s$-channel & $tW$-channel  
\\ 
\hline
8 TeV  & $86.5 {}^{+2.8}_{-1.0} {}^{+2.0}_{-2.2}$ & $5.65 \pm 0.08 \pm 0.21$
& $22.0 \pm 0.6 \pm 1.4$ 
\\ 
13 TeV & $218 {}^{+5}_{-2} \pm 5$ & $11.17 \pm 0.18 \pm 0.38$
& $70.4 \pm 1.8 {}^{+3.2}_{-3.4}$
\\ 
14 TeV & $248 {}^{+6}_{-2} {}^{+5}_{-6}$ & $12.35 \pm 0.19 {}^{+0.49}_{-0.41}$
& $83.1 \pm 2.0 {}^{+3.1}_{-4.6}$
\end{tabular}
\caption{aNNLO single-top total cross sections with $m_t=173.3$ GeV in the $t$, $s$, and $tW$ channels. The results include the sum of the single-top and single-antitop cross sections in each channel.}
\label{table3}
\end{center}
\end{table}

In Table \ref{table3} we show the sum of the single-top and single-antitop 
production cross sections at aNNLO at the LHC in the $t$, $s$, and $tW$ channels. The first uncertainty is from scale variation while the second is from pdf errors with MSTW2008 NNLO pdf at 90\% CL \cite{MSTW2008}. 

\begin{figure}
\begin{center}
\includegraphics[width=.65\textwidth]{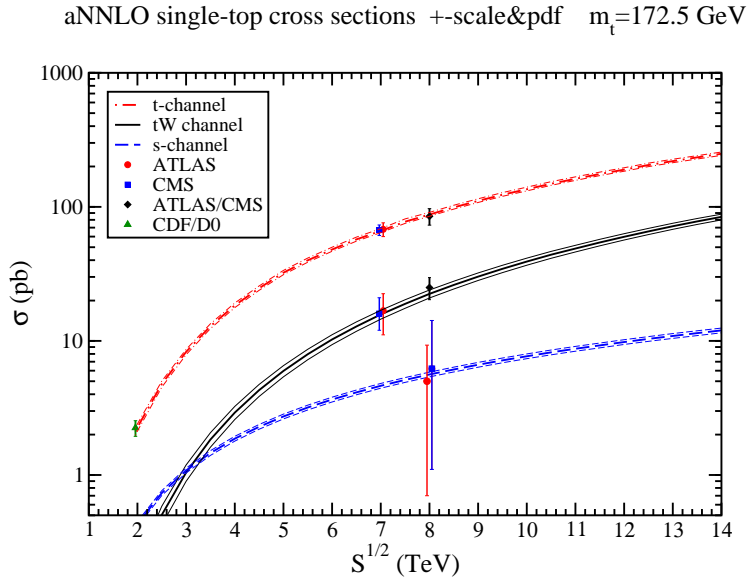}
\caption{Single-top cross sections in all three channels at LHC energies.}
\label{figsingletop}
\end{center}
\end{figure}

In Fig. \ref{figsingletop} we plot the total cross sections (again, sum of single-top and single-antitop) in each channel versus collider energy.  At 1.96 TeV energy we compare with $t$-channel CDF/D0 combination data \cite{tchtevcombo}. At 7 TeV LHC energy, we compare with $t$-channel data from ATLAS \cite{ATLAStch7lhc} and CMS \cite{CMStch7lhc}, and with $tW$ data from ATLAS \cite{ATLAStW7lhc} and CMS \cite{CMStW7lhc}.
At 8 TeV LHC energy we compare with ATLAS/CMS combination $t$-channel data \cite{tch8lhccombo}, with $s$-channel data from ATLAS \cite{ATLASsch8lhc} and CMS \cite{CMSsch8lhc}, and with $tW$-channel ATLAS/CMS combination data \cite{tW8lhccombo}. We find excellent agreement of theory with data for all collider energies in all channels.
We also note that the aNNLO $t$-channel ratio of top and antitop cross sections compares well with data as well as with the NNLO result in \cite{NNLOtch}. 

\begin{figure}
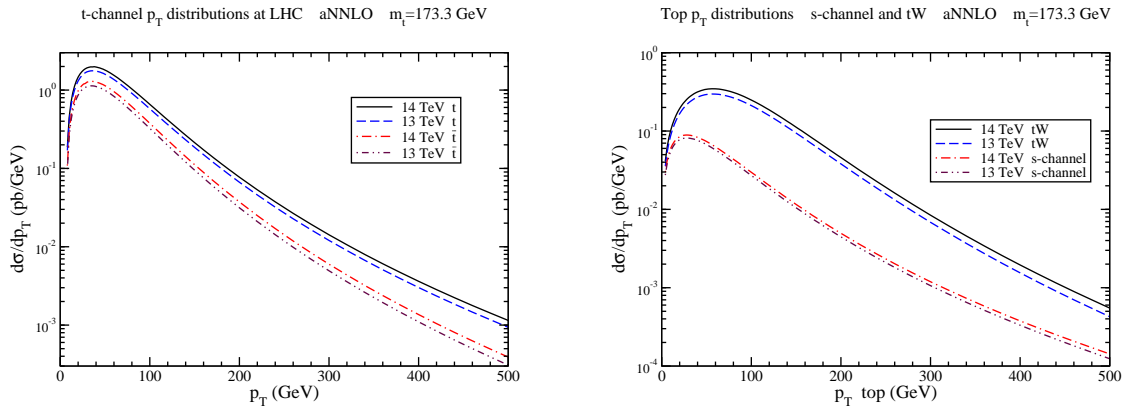

\begin{center}
\includegraphics[width=.45\textwidth]{pttch1314lhcplot.eps}
\hspace{10mm}
\includegraphics[width=.45\textwidth]{ptschtW1314lhcplot.eps}
\caption{$t$-channel (left), $s$-channel (right), and $tW$-channel (right) top-quark $p_T$ distributions at 13 and 14 TeV LHC energies.}
\label{tstWpt}
\end{center}
\end{figure}

We continue with $t$-channel aNNLO $p_T$ distributions \cite{NKtchpt}. In the left plot of Fig. \ref{tstWpt} we display the top and antitop $p_T$ distributions in $t$-channel production at 13 and 14 TeV LHC energies. We also note that the $t$-channel distributions at 7 TeV LHC energy compare well with ATLAS data \cite{ATLAStch7lhc} as shown in \cite{NK15tWZ}.
Top $p_T$ distributions in $s$-channel and in $tW$ production at 13 and 14 TeV LHC energies are displayed in the right plot of Fig. \ref{tstWpt}.

\section{Top production via anomalous gluon couplings}

Finally, we discuss top-quark production in processes with anomalous gluon couplings. The partonic processes are of the form $gu \rightarrow tg$ which involve 
$t$-$u$-$g$ couplings. We studied these processes and calculated the soft-gluon corrections at NLL accuracy in \cite{NKEM}. 

The soft-gluon corrections are large and they reduce the scale dependence of the cross section. At both 7 and 14 TeV energies, the NLO soft-gluon corrections increase the LO cross section by around 60\% for $\mu=m_t$. The reduction in scale variation over $m_t/2 \le \mu \le 2m_t$ is also very significant.

\section{Summary}

The soft-gluon corrections from NNLL resummation for top-pair production have been calculated at N$^3$LO. The corrections are sizable 
and they provide the best available theoretical predictions.
Results have been presented for the total $t{\bar t}$ cross sections,
the top quark $p_T$ and rapidity distributions, and the top quark 
forward-backward asymmetry.
The corrections are large for LHC and Tevatron energies and they reduce the 
theoretical uncertainties from scale variation.
There is excellent agreement between theoretical predictions and LHC and 
Tevatron data.

We have also presented results for NNLL soft-gluon corrections at NNLO in single-top production. The corrections are significant at the LHC and the Tevatron. 
There is excellent agreement with LHC and Tevatron data in all single-top channels. We also discussed single-top production via anomalous gluon couplings.

\end{document}